\documentclass[pra,aps,showpacs,twocolumn]{revtex4}
\usepackage{epsfig}
\usepackage{color}

\pacs{03.75.Kk,03.75.Lm,67.85.Jk}  

\begin{document} 
\title{Coherence and Josephson oscillations 
between two tunnel-coupled one-dimensional atomic quasicondensates at finite temperature} 

\author{Pjotrs Gri\v{s}ins$^1$ and Igor E. Mazets$^{1,2}$}
\affiliation{$^1$Vienna Center for Quantum Science and Technology, Atominstitut, TU Wien, 1020 Vienna, Austria \\
$^2$Ioffe Physico-Technical Institute of the Russian Academy of Sciences, 194021 St.Petersburg, Russia } 

\begin{abstract} 
We revisit the theory of tunnel-coupled atomic quasicondensates in double-well elongated traps at finite 
temperatures. Using the 
functional integral approach, we calculate the relative phase correlation function beyond the harmonic limit of small  
fluctuations of the relative phase and its conjugate relative-density variable. 
We show that the thermal fluctuations of the relative 
phase between the two quasicondensates  decrease  the frequency of 
Josephson oscillations and even wash out these oscillations for small values of the tunnel coupling. 
\end{abstract} 

\maketitle

\section{Introduction} 
\label{sec.I}

Systems of ultracold bosonic atoms in two parallel atomic waveguides mutually coupled via quantum tunneling  
(so-called extended bosonic Josephson junctions) have been a subject of intensive theoretical 
\cite{WB,Bd,LvK,HP,Stim1,Brand,MSLF} and experimental \cite{Betz} studies. The finite spatial extension 
of these systems provides much richer physics compared to the case of a point-like bosonic Josephson 
junction \cite{Smerzi}. The novel features arise due to the enhanced role of noise and correlations in 
low-dimensional ultracold atomic systems. 

Before discussing the effects of tunneling, we recall the basic properties of a bosonic system in an 
isolated waveguide \cite{Popov,MoC,Caza}. 
This system is effectively one-dimensional (1D), if the interaction energy per atom 
(we assume interatomic repulsion characterized by the effective 1D coupling strength $g>0$) and the 
temperature are well below the spacing between the discrete energy levels of the potential of tight radial 
confinement. In this case quantum degeneracy does not lead to establishment of the  long-range order; instead, 
atoms form a quasicondensate, i.e. a system describable by a macroscopic wave function with strong phase 
fluctuations. The characteristic length of the phase coherence in a quasicondensate at finite temperature $T$ is 
$\lambda _T=2\hbar ^2n_\mathrm{1D}/(mk_\mathrm{B}T)$, where $m$ is the atomic mass, and 
$n_\mathrm{1D}$ is the mean  linear density of atoms \cite{Popov} 
(we assume an infinite system; thermodynamic limit implies constant $n_\mathrm{1D}=N/L$ while both the atom number $N$ 
and the quantization length $L$ tend to infinity). The power-law decrease of the single-particle correlation function 
takes place only at $T=0$. 

If  two waveguides  are tunnel-coupled,  the system is 
described by the generalized Hamiltonian 
\begin{eqnarray} 
\hat{H}&=&\int dz\, \Bigg {[} \sum _{j=1}^2 \Bigg {(} \frac {\hbar ^2}{2m} \frac {\partial \hat{\psi }^\dag _j}{\partial z} 
\frac {\partial \hat{\psi }_j}{\partial z} +\frac {g}2 \hat{\psi }^\dag _j\hat{\psi }^\dag _j\hat{\psi }_j\hat{\psi }_j - 
\nonumber \\ && 
-\mu \hat{\psi }^\dag _j\hat{\psi }_j \Bigg {)} 
-\hbar J \left( \hat{\psi }^\dag _1 \hat{\psi }_2+\hat{\psi }^\dag _2\hat{\psi }_1\right) \Bigg {]}, 
\label{Hmlt} 
\end{eqnarray}  
where $\hat{\psi }_j$ is the atomic annihilation operator for the $j$th waveguide ($j=1,2$), 
$\mu =\hbar gn_\mathrm{1D} -\hbar J$ is the chemical potential and $2J$ is the tunnel splitting (in frequency units),  
i.e., the frequency interval between the two lowest eigenstates of the radial trapping  Hamiltonian 
(the antisymmetric and symmetric superpositions of the single-atom states localized in either $j=1$ or $j=2$ wells of 
the double-well Hamiltonian). In this case 
the situation changes qualitatively: the tunnel coupling mutually locks phase fluctuations in the two quasicondensates \cite{WB}. 
Phase locking (as we shall quantify later, in Sec. \ref{sec.II}) means that the distribution of the relative phase between the two quasicondensates becomes 
peaked around zero, while the local phase of an individual ($j=1$ or 2) quasicondensate remains fully random (the phase-density representation for 
quasicondensates will be discussed in Sec. \ref{sec.II}). 
In the spatial correlation of the local relative phase between two quasicondensates a new length parameter 
appears \cite{WB,Stim1}, 
\begin{equation} 
l_J = \sqrt{\hbar /(4mJ)} .
\label{mlength} 
\end{equation} 
The length $l_J$ sets the scale of restoration of the inter-waveguide 
coherence due to finite tunnel-coupling strength $J$. The tunnel-coupling 
strength is usually estimated from the single-particle energy (kinetic and potential) 
and the overlap in the potential barrier region  
of the wave functions for a particle localized in the 1st and 2nd waveguide. However, 
it is also possible to take into account atomic interactions, see Ref. \cite{Brand} and references therein. 

Experimentally, the interwell coherence can be observed by releasing the two quasicondensates from the trap and measuring 
locally the contrast and the phase of their interference pattern after time of flight \cite{H1,H2}. 

Up to now, only the theory based on linearization of the Hamiltonian (\ref{Hmlt}) has been developed \cite{WB} and 
applied to the analysis of the experimental data \cite{Stim1,Betz}. Our work is aimed to develop a model of the steady-state thermal 
noise in tunnel-coupled quasicondensates beyond the harmonic approximation as well as to quantify the influence of the thermal 
noise to the macroscopic coherent dynamics of the system (Josephson oscillations). 

Our paper is  organized as follows. In Sec. \ref{sec.II} we summarize the harmonic approach of Ref. \cite{WB}. Section \ref{stat} is divided 
in two Subsections. In Subsection \ref{subIII.1} 
we calculate the static correlation properties of our system beyond the harmonic approximation using the transfer operator 
technique in the classical limit. The condition for negligibility of the quantum noise is also derived. Our way to model equilibrium state 
by numerical simulation of the system's relaxation to the equilibrium after a quench is explained in Subsection \ref{subIII.2}. 
Section \ref{jxq} deals with the noise-affected Josephson oscillations. We 
derive analitically the frequency of Josephson oscillations modified by the thermal noise in our extended system. 
We support our analysis by
numerical simulations and also observe Josephson oscillations
washing out as a result of thermal noise for low enough
tunnel coupling. Section \ref{concl} contains our final remarks and conclusions. 
Explanations of the ways to derive the main equations of Sec. \ref{stat} and Sec. \ref{jxq} 
are presented in Appendices \ref{A.A} and \ref{A.B}, respectively.

\section{Harmonic approximation}
\label{sec.II} 

Following the standard procedure \cite{MoC}, 
we represent our atomic field operators through the phase $\hat \theta _j(z)$ and density $\hat \rho _j(z)$ operators, obeying the 
commutation relation $[\hat \theta _j(z),\hat \rho _{j^\prime }(z^\prime )]=-i\delta (z-z^\prime )\delta _{j\, j^\prime }$, as 
\begin{equation} 
\hat \psi _j (z)=\exp [i\hat \theta _j(z)] \sqrt{ \hat \rho _j(z)} ,\qquad j=1,2. 
\label{ph-d-repr} 
\end{equation}  
A discussion of the way to introduce the phase operator for quasicondensates by coarse graining a lattice model  
on  length scales containing sufficiently many atoms can be found in Ref. \cite{MoC}. 
The density operator can be represented as $\hat \rho _j(z) =n_\mathrm{1D}+ 
\delta \! \hat \rho _j(z)$. Since for quantum gases with repulsive atomic interactions  density fluctuations are suppressed, we can 
always consider the corresponding operator $\delta \! \hat \rho _j$ as a small correction. However, the same is not always true for the 
phase fluctuations. 

Whitlock and Bouchoule \cite{WB} from the very beginning assumed the phase fluctuations to be small   
and thus linearized the Hamiltonian Eq. (\ref{Hmlt}) 
reducing it to $\hat{H}\approx \hat{H}_\mathrm{lin}$, 
\begin{eqnarray} 
\hat{H}_\mathrm{lin}&=&\int dz \Bigg{[} \frac {\hbar ^2n_\mathrm{1D}}m \left( \frac {\partial \hat{ \theta } _\mathrm{s}}{\partial z}\right) ^2+
\frac {\hbar ^2}{16mn_\mathrm{1D}} \left( \frac {\partial \delta \! \hat{\rho }_\mathrm{s}}{\partial z}\right) ^2+ \nonumber \\ &&
\frac { g}4\delta \! \hat{\rho }_\mathrm{s}^2 + \frac {\hbar ^2n_\mathrm{1D}}{4m} \left( \frac {\partial \hat{\theta }_\mathrm{a}}{\partial z}\right) ^2+
\frac {\hbar ^2}{4mn_\mathrm{1D}} \left( \frac {\partial \delta \! \hat{\rho }_\mathrm{a}}{\partial z}\right) ^2+ \nonumber \\ && 
{g}\, \delta \! \hat{\rho }_\mathrm{a}^2 +{\hbar Jn_\mathrm{1D}}\hat{ \theta }_\mathrm{a}^2\Bigg{]} .
\label{harmquant} 
\end{eqnarray}
Here the symmetric (s) and antisymmetric (a) variables are introduced via canonical transformation 
\begin{eqnarray} 
\delta \! \hat \rho _\mathrm{s}(z)&=& \delta \! \hat \rho _1 (z)+ \delta \! \hat \rho _2(z), ~~
\hat \theta _\mathrm{s} (z)=[\hat \theta _1 (z)+\hat \theta _2 (z)]/2, \nonumber \\ 
\delta \! \hat \rho _\mathrm{a}(z)&=& [\delta \! \hat \rho _1 (z)- \delta \! \hat \rho _2(z)]/2, ~~ 
\hat \theta _\mathrm{a} (z)=\hat \theta _1 (z)-\hat \theta _2 (z). \nonumber  
\end{eqnarray} 

Diagonalization of the Hamiltonian (\ref{harmquant}) is based on the 
Fourier transform   $\delta \! \hat \rho _\mathrm{s(a)} (z)=L^{-1/2}\sum _{k\neq 0} 
\delta \!  \hat \rho _{\mathrm{s(a)},k} e^{ikz}$,  $\hat \theta _\mathrm{s(a)} (z)=L^{-1/2}\sum _{k\neq 0} 
\hat \theta _{\mathrm{s(a)},k} e^{ikz}$. The frequencies $\omega _\mathrm{s(a)}(k)$ of the symmetric and antisymmetric modes with the momentum $\hbar k$ are given 
by  the dispersion relations  
\begin{eqnarray} 
\omega _\mathrm{s}^2(k)&=&\frac {\hbar k^2}{2m} \left( \frac {\hbar k^2}{2m}+\frac {2gn_\mathrm{1D}}\hbar \right) , \label{disp.law.s} \\ 
\omega _\mathrm{a}^2(k)&=&\left( \frac {\hbar k^2}{2m}+2J\right)  \left( \frac {\hbar k^2}{2m}+2J+\frac {2gn_\mathrm{1D}}\hbar \right) . 
\label{disp.law.a}
\end{eqnarray} 

Correlations in two tunnel-coupled quasicondensates are experimentally accessible via the two-point correlation function 
$g^\mathrm{a}_2(z-z^\prime ) =n_\mathrm{1D}^{-2}\langle :\!\hat{\psi }^\dag _1(z)\hat{\psi }_2^\dag (z^\prime )\hat{\psi }_2(z)\hat{\psi }_1 (z^\prime )\!:\rangle $. 
Since the system described by the Hamiltonian (\ref{Hmlt}) is translationally invariant, $g^\mathrm{a}_2$ depends only on the difference of the two 
co-ordinates. The symbol $\langle : \hat{O} :\rangle $ denotes the average of the normal ordered (with respect to the atomic operators $\hat \psi _j$, 
$\hat \psi _j^\dag $) form of the operator $\hat O$. In what follows, we omit the normal ordering notation, thus neglecting the atomic shot noise. 

Since the density fluctuations for $|k|\lesssim \xi ^{-1}$, $\xi =\hbar / \sqrt{mgn_\mathrm{1D}}=\hbar /(mc)$ being the healing length, are 
suppressed by the atomic repulsion \cite{Popov,MoC}, the main contribution to this correlation function is given by the phase fluctuations, 
$g^\mathrm{a}_2(z-z^\prime )\approx  \langle \exp [ i\hat \theta _\mathrm{a}(z^\prime ) -
i\hat \theta _\mathrm{a}(z)]\rangle $. 

The experimentally accessible length scale cannot be shorter than the optical resolution length $\Delta z_\mathrm{opt}$. On this scale the shot noise 
yields the quantum uncertainty of the relative phase, coarse grained over the distance $\Delta z_\mathrm{opt}$, of the order of 
$1/\sqrt{2n_\mathrm{1D}\Delta z_\mathrm{opt}}$. For $\Delta z_\mathrm{opt}\gtrsim 3 ~\mu \mathrm{m}$ and $n_\mathrm{1D}\gtrsim 30 ~\mu \mathrm{m}^{-1}$ 
the shot-noise induced phase uncertainty does not exceed 0.075~rad. This relatively small value  can be always kept in mind when comparing 
theoretical predictions to measurement results. However, for the sake of simplicity, in what follows we assume  
$\langle :\exp [ i\hat \theta _\mathrm{a}(z^\prime ) -
i\hat \theta _\mathrm{a}(z)]:\rangle \approx 
\langle \exp [ i\hat \theta _\mathrm{a}(z^\prime ) -
i\hat \theta _\mathrm{a}(z)]\rangle $ and so on. 

Another point related to the use of the fully classical approximation is the substitution of the Bose-Einstein statistics of the elementary excitations 
by its classical limit, 
\begin{equation} 
\frac 1{\exp [\hbar \omega _\mathrm{a}(k)/(k_\mathrm{B}T)]-1}\approx  \frac {k_\mathrm{B}T}{\hbar \omega _\mathrm{a}(k)}. 
\label{eq.new5} 
\end{equation} 
One obtains strong deviations from Eq. (\ref{eq.new5}) for $\hbar \omega _\mathrm{a}(k)\gtrsim k_\mathrm{B}T$, which corresponds, under typical experimental 
conditions, to the range of wave lengths shorter than $\Delta z_\mathrm{opt}$, i.e., not resolvable optically. 

These considerations justify our method based on genuinely classical statistics. 

In the harmonic approximations fluctuations are Gaussian, hence, $\langle \exp [ i\hat \theta _\mathrm{a}(z^\prime ) -
i\hat \theta _\mathrm{a}(z)]\rangle = \exp \{ -\frac 12 \langle [ \hat \theta _\mathrm{a}(z^\prime ) -
\hat \theta _\mathrm{a}(z)]^2\rangle \} $. Expressing $\hat \theta _\mathrm{a}$ through creation and annihilation operators of the elementary 
excitations and calculating thermal populations of the elementary modes using Eq. (\ref{eq.new5}), Whitlock and Bouchoule obtained \cite{WB} 
\begin{equation} 
\langle \exp [ i\hat \theta _\mathrm{a}(z^\prime ) -
i\hat \theta _\mathrm{a}(z)]\rangle = 
\exp \left[ -\frac {2l_J}{\lambda _T}(1-e^{-|z-z^\prime |/l_J})\right] .
\label{cfB} 
\end{equation} 
From this expression we can see that tunnel coupling locks the relative phase between two quasicondensates. This locking means that 
the relative-phase correlation function (\ref{cfB}) does not decrease to zero, but even at $ |z-z^\prime |\rightarrow \infty $ has a 
finite value, corresponding to $\langle \hat \theta _\mathrm{a}^2(z) \rangle = {2l_J}/{\lambda _T}$. On the contrary, the phase correlations in each of the 
waveguides are $\langle \exp [ i\hat \theta _j(z^\prime ) -
i\hat \theta _j(z)]\rangle = \langle \exp \{ i[\hat \theta _\mathrm{s}(z^\prime )\pm \frac 12\hat \theta _\mathrm{a}(z^\prime ) -
\hat \theta _\mathrm{s}(z )\mp \frac 12\hat \theta _\mathrm{a}(z)]\} \rangle $, the  upper and lower signs corresponding to 
$j=1$ and $j=2$, respectively. We can evaluate them using the statistical independence of noise in the symmetric and antisymmetric modes. The result  
\begin{eqnarray} 
&&\! \! \! \! \langle \exp [ i\hat \theta _j(z^\prime ) -
i\hat \theta _j(z)]\rangle = \exp \Bigg{ \{ }-\frac 12 \langle [\hat \theta _\mathrm{s}(z^\prime )-\hat \theta _\mathrm{s}(z )]^2\rangle - \nonumber \\ &&
\qquad \qquad \qquad \qquad \qquad \qquad \quad \frac 18 \langle [\hat \theta _\mathrm{a}(z^\prime )-\hat \theta _\mathrm{a}(z)]^2\rangle \Bigg{ \} }\nonumber \\ 
&&\qquad = \exp \Bigg{[} - \frac {|z-z^\prime |}{2\lambda _T}-   
\frac {l_J}{2\lambda _T}(1-e^{-|z-z^\prime |/l_J})\Bigg{]}   
\label{eq.new6}  
\end{eqnarray}  
decreases $\propto \exp [ -|z-z^\prime |/(2\lambda _T)] $ at 
$ |z-z^\prime |\rightarrow \infty $ because of the unlimited growth  of the fluctuations of the symmetric component of the phase 
along the $z$-direction. 
The correlation properties of the symmetric mode can be experimentally measured using the density-density correlations of the ultracold gas 
in a time-of-flight experiment \cite{Manz1}, however, this subject is beyond the scope of our present paper.  

The phase locking of the relative phase becomes most apparent if we treat the evolution of the relative phase along $z$ in the harmonic approximation 
as the Ornstein-Uhlenbeck stochastic process \cite{Stim1}: while thermal excitations result in the relative phase diffusion, with the diffusion 
coefficient proportional to $\lambda _T^{-1}$, the tunnel coupling gives rise to the ``friction" force that tends to restore a small (ultimately zero) 
local phase difference between the two quasicondensates.

\section{Correlation functions and the interwell coherence beyond the harmonic approximation}  
\label{stat} 

\subsection{Equilibrium theory} 
\label{subIII.1}

In the present work we make a step further with respect to the theory of Ref. \cite{WB}  
and abandon the assumption of small phase fluctuations (but still consider small 
density fluctuations, which is a reasonable approximation for quasicondensates with repulsive interactions). We evaluate the partition function \cite{Popov}  
\begin{equation} 
Z=\int {\cal D}\delta \! \rho _\mathrm{s}\int {\cal D}\theta _\mathrm{s} \int {\cal D}\delta \! \rho _\mathrm{a}\int {\cal D}\theta _\mathrm{a}
\, \exp [-H/(k_\mathrm{B}T)] , 
\label{eqZ1} 
\end{equation}  
where 
\begin{eqnarray} 
H&=&\int dz \left[ \frac {\hbar ^2n_\mathrm{1D}}m \left( \frac {\partial \theta _\mathrm{s}}{\partial z}\right) ^2+
\frac { g}4\delta \! \rho _\mathrm{s}^2 +\right. \nonumber \\ &&
\left. \frac {\hbar ^2n_\mathrm{1D}}{4m} \left( \frac {\partial \theta _\mathrm{a}}{\partial z}\right) ^2+
{g}\, \delta \! \rho _\mathrm{a}^2 +{2\hbar Jn_\mathrm{1D}}(1-\cos \theta _\mathrm{a})\right] ~~~~
\label{Hfullnew} 
\end{eqnarray}  
is the Hamiltonian (\ref{Hmlt}) expressed through the classical fields $\delta \! \rho _\mathrm{s,a},\, \theta _\mathrm{s,a}$ 
(in the co-ordinate representation), over 
which the functional integrals are taken. For the sake of simplicity, we write the Hamiltonian (\ref{Hfullnew}) in the 
phononic limit, where the fluctuation wavelengths are long compared to the healing length of the quasicondensate and Eqs. 
(\ref{disp.law.s},~\ref{disp.law.a}) are reduced to $\omega _\mathrm{s} ^2(k)\approx c^2k^2$ and 
$\omega _\mathrm{a} ^2(k)\approx c^2k^2+4Jgn_\mathrm{1D}/\hbar $, where $c=\sqrt{gn_\mathrm{1D}/m}$ is the speed of sound. 
Of course, the phase-density description can be extended into short-wavelength excitation range \cite{Popov,MoC}, bringing 
about the Hamltonian terms $\propto (\partial \delta \! \rho _\mathrm{s,a}/\partial z)^2$ and 
thus revealing the full Bogoliubov-like spectra (\ref{disp.law.s},~\ref{disp.law.a}). However, we are not interested in 
the short-wavelength limit, since the respective length scales cannot be resolved by optical imaging systems 
\cite{Betz,H1,H2}. The system's description by 
Eq. (\ref{Hfullnew}) is fully consistent with Haldane's bosonization method \cite{Haldane}. The relative phase $\theta _\mathrm{a}$ 
is accessible through interference patterns observed in time-of-flight experiments \cite{Betz,H1,H2}. We develop here the way to 
evaluate its correlation properties. 
Since the density fluctuations are small, we can  decouple symmetric and antisymmetric modes \cite{GPD} and integrate out the 
variables of the symmetric mode. The absence of cross-terms containing both $\delta \! \rho _\mathrm{a}$ and 
$\theta _\mathrm{a}$ in Eq. (\ref{Hfullnew}) allows us to 
integrate out $\delta \! \rho _\mathrm{a}$ as well and to obtain, as an intermediate result, the partition function in the form 
\begin{eqnarray}  
Z&=&\mathrm{const} \, \int {\cal D}\theta _\mathrm{a}\, \exp \Bigg {\{ }-\int dz \, 
\Bigg {[} \frac {\hbar ^2 n_\mathrm{1D}}{4mk_\mathrm{B}T} 
\left( \frac {\partial \theta _\mathrm{a}}{\partial z}\right) ^2  + \nonumber \\ && 
\frac {2\hbar Jn_\mathrm{1D}}{k_\mathrm{B}T}(1-\cos \theta _\mathrm{a})\Bigg {]} \Bigg {\} }  
\label{eqZ2} 
\end{eqnarray} 
that was considered long ago \cite{Kr1,Kr2} in the context of the statistical mechanics of systems describable by the sine-Gordon equation, 
which is known to adequately account for the low-energy physics of tunnel-coupled 1D ultracold atomic systems \cite{GPD}. 

Note that anharmonic Hamiltonian terms, which depend on the density fluctuations neglected in our present theory, do not affect much 
the static properties of the quasicondensate \cite{MoC}. One needs to take them into account in the analysis \cite{Stimming23} of a slow process of 
the system's relaxation towards equilibrium starting from  
a non-equilibrium, pre-thermalized initial state \cite{Gring}, characterized by two different temperatures $T_+$ and $T_-\ll T_+$ 
for the symmetric and antisymmetric modes, respectively. 

The applicability range of our fully classical approach can be determined as follows. Consider, for the sake of simplicity, distances 
shorter than $l_J$. The effects of tunnel coupling can be neglected at such short length scales, and the fully classical 
correlation function can be estimated \cite{WB} as $\langle \exp [ i \theta _\mathrm{a}(z^\prime ) -
i\theta _\mathrm{a}(z)]\rangle \approx \exp (-2|z-z^\prime |/\lambda _T)$. We have to compare this result to the power-law decay of 
correlations due to \textit{quantum} effects, which is obtained in the limit $T\rightarrow 0$ \cite{Popov,MoC}. Neglecting, as 
previously, the 
contribution of the density fluctuations, we can write $\lim _{T\rightarrow 0} \langle \exp [ i \hat{\theta }_\mathrm{a}(z^\prime ) -
i\hat{\theta }_\mathrm{a}(z)]\rangle \approx  
\lim _{T\rightarrow 0} \langle \hat{\psi }^\dag _1 (z^\prime )\hat{\psi }_1 (z) \rangle 
\langle \hat{\psi }^\dag _2 (z )\hat{\psi }_2 (z^\prime ) \rangle $ and, finally, 
\begin{equation} 
\lim _{T\rightarrow 0} \langle \exp [ i \hat{\theta }_\mathrm{a}(z^\prime ) -
i\hat{\theta }_\mathrm{a}(z)]\rangle \approx \left( \frac {\Lambda _\mathrm{UV}}{|z-z^\prime |} \right) ^{1/{\cal K}},  
\label{eq.new1} 
\end{equation} 
where the quantum mechanical average over the ground state is taken, ${\cal K}= \pi \hbar \sqrt{n_\mathrm{1D}/(mg)}$ is the Luttinger liquid 
parameter (for quasicondensates, which are weakly interacting systems, ${\cal K}\gg 1$), and $\Lambda _\mathrm{UV} $ is the ultraviolet cutoff 
of the theory. Eq. (\ref{eq.new1}) is valid if  
\begin{equation} 
|z-z^\prime |\gg \Lambda _\mathrm{UV} . 
\label{eq.new2} 
\end{equation} 
The estimation by Popov \cite{cutoff} yields $\Lambda _\mathrm{UV} \sim \xi $. 

We can fully neglect quantum fluctuations if their contribution to the decay of correlations is small, compared to the contribution of the 
thermal noise, on a given length scale. 
The correlation decay is dominated by the thermal noise if the classical formula $\exp (-2|z-z^\prime |/\lambda _T )$ 
yields stronger decay of correlations than the quantum limit (\ref{eq.new2}), i.e., if  
\begin{equation} 
2|z-z^\prime |/\lambda _T  \gtrsim {\cal K}^{-1}\ln  \left( {|z-z^\prime |}/\xi  \right) .
\label{eq.new3} 
\end{equation} 
The experimentally 
relevant range of $|z-z^\prime |$ is bound from below by  $\Delta z_\mathrm{opt}$, as we discussed in Sec. \ref{sec.II}, 
and $\Delta z_\mathrm{opt}\gg \xi $ in a typical experiment \cite{Betz}. Therefore the use of the fully classical approach is reasonable for 
\begin{equation} 
k_\mathrm{B}T \gtrsim mc^2 \frac {\xi \ln (\Delta z_\mathrm{opt}/\xi )}{\pi \Delta z_\mathrm{opt}} . 
\label{eq.new4} 
\end{equation}

We can evaluate the partition function (\ref{eqZ2}) using the transfer operator technique \cite{Kr1,Kr2,transfer1}. First of all, we evaluate the 
phase-correlation function as (see Appendix \ref{A.A} for the sketch of derivation) 
\begin{eqnarray} 
\langle \exp [ i \theta _\mathrm{a}(z^\prime ) -
i\theta _\mathrm{a}(z)]\rangle =\qquad \qquad \qquad \qquad  && \nonumber \\ 
\sum _{n=0}^\infty \left| \langle n |e^{i\theta }|0\rangle \right| ^2 \exp [ -(\epsilon _n-\epsilon _0) 
|z-z^\prime | ] &,&  
\label{cfG} 
\end{eqnarray} 
where 
\begin{equation} 
\langle n |e^{i\theta }|0\rangle =\int _{-\pi}^\pi d\theta \, \Psi _n^*(\theta )e^{i\theta }\Psi _0(\theta ), 
\label{aux1} 
\end{equation} 
$\Psi _n(\theta )$ is the eigenfunction (normalized to 1) of the auxiliary Schr\"odinger-type equation 
\begin{equation}  
\left[ -\frac {2}{\lambda _T} \frac {\partial ^2}{\partial \theta ^2} - \frac {\lambda _T}{4l_J^2}
(\cos \theta -1)\right] \Psi _n(\theta )=\epsilon _n\Psi _n(\theta ) ,
\label{aux2} 
\end{equation} 
and $\epsilon _n$, $n=0,1,2, \dots $, is the respective eigenvalue. For simplicity, we set  periodic 
(and not quasiperiodic) boundary conditions to Eq. (\ref{aux2}) with 
the period $2\pi $, thus neglecting the band structure of its spectrum, since the zero-quasimomentum solutions define all the system properties \cite{Kr2}, 
which are relevant to our present work. 

In the limit of strong tunnel coupling, $l_J\ll \lambda _T$, the operator in the left-hand-side of Eq. (\ref{aux2}) can be approximated by the harmonic oscillator 
Hamiltonian (in proper units), and $\epsilon _n =l_J^{-1}(n+\frac 12)$, $n=0,1,2, \dots ~ $. In this limit Eq. (\ref{cfG}) 
reproduces the result (\ref{cfB}) that holds for small phase fluctuations. 

In the opposite limit, Eq.~(\ref{aux2}) can be solved perturbatively, and we obtain 
\begin{eqnarray}  
\langle \exp [ i  \theta _\mathrm{a}(z^\prime ) -
i  \theta _\mathrm{a}(z)]\rangle \approx \left( \frac {\lambda _T^2}{8l_J^2}\right) ^2 + \qquad \qquad \qquad && \nonumber \\ 
\left[ 1- \left( \frac {\lambda _T^2}{8l_J^2}\right) ^2
\right] \exp \left( -\frac {2|z-z^\prime |}{\lambda _T} \right) , \quad  {l_J}\gg {\lambda _T}. 
\label{highT} 
\end{eqnarray}  

In what follows, we will be interested in calculating the value of 
\begin{equation} 
\langle \cos  \theta _\mathrm{a} \rangle = \langle 0| \cos \theta |0\rangle , 
\label{cosgen} 
\end{equation} 
which can be viewed as the mean interwell coherence. 
This expression can be derived in different ways, e.g., from Eq. (\ref{cfG}) by employing the statistical independence 
of phase fluctuations at two very distant points, $|z-z^\prime |\rightarrow \infty $, and recalling that $\langle \sin \hat \theta _\mathrm{a}\rangle =0$.  
In a general case, Eq. (\ref{cosgen}) can be evaluated from the lowest-energy solution of the Mathieu equation \cite{Abr}. In the two limiting cases we obtain 
the asymptotics 
\begin{equation}  
\langle \cos  \theta _\mathrm{a} \rangle \approx \left \{ 
\begin{array}{ll}
\exp (-l_J/\lambda _T), & l_J\ll \lambda _T \\
\lambda _T^2/(8 l_J^2), & l_J\gg \lambda _T 
\end{array} 
\right.  . 
\label{asc}
\end{equation} 
A possible physical explanation of the fact that the mean interwell coherence decreases at $l_J/\lambda _T\rightarrow \infty $ much slower 
that the harmonic approximation \cite{WB} predicts, 
is the large probability of thermal excitation of a soliton in this limit. Each emerging soliton decreases the 
number of phononic states by 1 \cite{Kr2}, and the phononic density of states is reduced mostly in the long-wavelength range (for phonon momenta less than or  
of the order of $\hbar /l_J$), which gives the main contribution to the long-distance behavior of the correlation function (\ref{cfG}) and, hence, to 
$\langle \cos  \theta _\mathrm{a} \rangle $. 

\subsection{Relaxation to the equilibrium after a quench}
\label{subIII.2} 

The results of Sec. \ref{subIII.1} are obtained at the equilibrium. However, it is interesting to investigate also 
the process of equilibration in the system of two 1D quasicondensates after a quench. The study of this dynamical 
problem is motivated by our recent numerical results \cite{Grisins} related to thermalization in a single 1D quasicondensate. 
In Ref. \cite{Grisins} we found that, despite the numerically confirmed 
integrability of the system, phononic (low-momentum) modes rapidly relaxed from their 
initial non-equilibrium state towards a final equilibrium state; particle-like (large-momentum) excitations, on the contrary,  
exhibited almost no relaxation. The equilibrium ensemble of phonons was different from the classical limit of 
equipartition of the thermal energy between all the degrees of freedom and was quite close to the Bose-Einstein 
distribution with the temperature $T_\mathrm{eff}$ 
determined by the total excitation energy of the initial non-equilibrium state. 
Observed fluctuations around this equilibrium state 
were  due to the finite size of the system inherent to numerical modeling. Remarkably, 
the correlations observed at the length scales, which are large compared to
the healing length to the healing length, as well as to the wavelength of an 
elementary excitation  with the energy equal to $k_\mathrm{B}T_\mathrm{eff}$, were well described by classical 
expressions. Note that the main contribution to the noise on these length scales stems from the low-energy 
excitations, which approximately exhibit  classical equipartition of energy. 

The need to extend  the numerical approach of Ref. \cite{Grisins} 
to tunnel-coupled 1D quasicondensates can also be seen from 
the following considerations. Our aim
is to numerically check the theoretically predicted correlations 
of two tunnel-coupled quasicondensates at equilibrium. This equilibrium state 
can be viewed as a result of the system's relaxation from its initial 
non-equilibrium state. Moreover, the available analytic theory predicts
only averages; unlike the case of harmonic approximation, there is no way yet
to generate individual realizations of the phase, obeying the necessary
statistics, without simulating numerically the equilibration process.
The most obvious way to obtain numerically the equilibrium solution is to
observe the numerical relaxation after a quench and wait until a steady-state
regime establishes. Particular type of the quench and the corresponding 
initial conditions are, up to a certain degree, arbitrary, as long as the 
system exhibits true relaxational dynamics.

Motivated by these considerations, 
we performed numerical modeling of the thermal equilibrium values of $\langle \cos \hat \theta _\mathrm{a} \rangle $ 
after the dynamical process of relaxation in our system after a quench. We simulated 
the time evolution of two coupled Gross-Pitaevskii equations using the split-step method \cite{num1} previously used by us \cite{Grisins} 
to simulate the dynamics of a single quasicondensate and now extended to the case of tunnel-coupled systems. 
As the initial conditions we took two independent quasicondensates with phonon modes populated randomly according to the 
Bose-Einstein thermal distribution. At $t=0$ we quenched the system by switching on 
the tunnel coupling between them. We solved this coupled system for a time long enough to provide equilibration. 

\begin{figure}[t]

\centerline{\epsfig{file=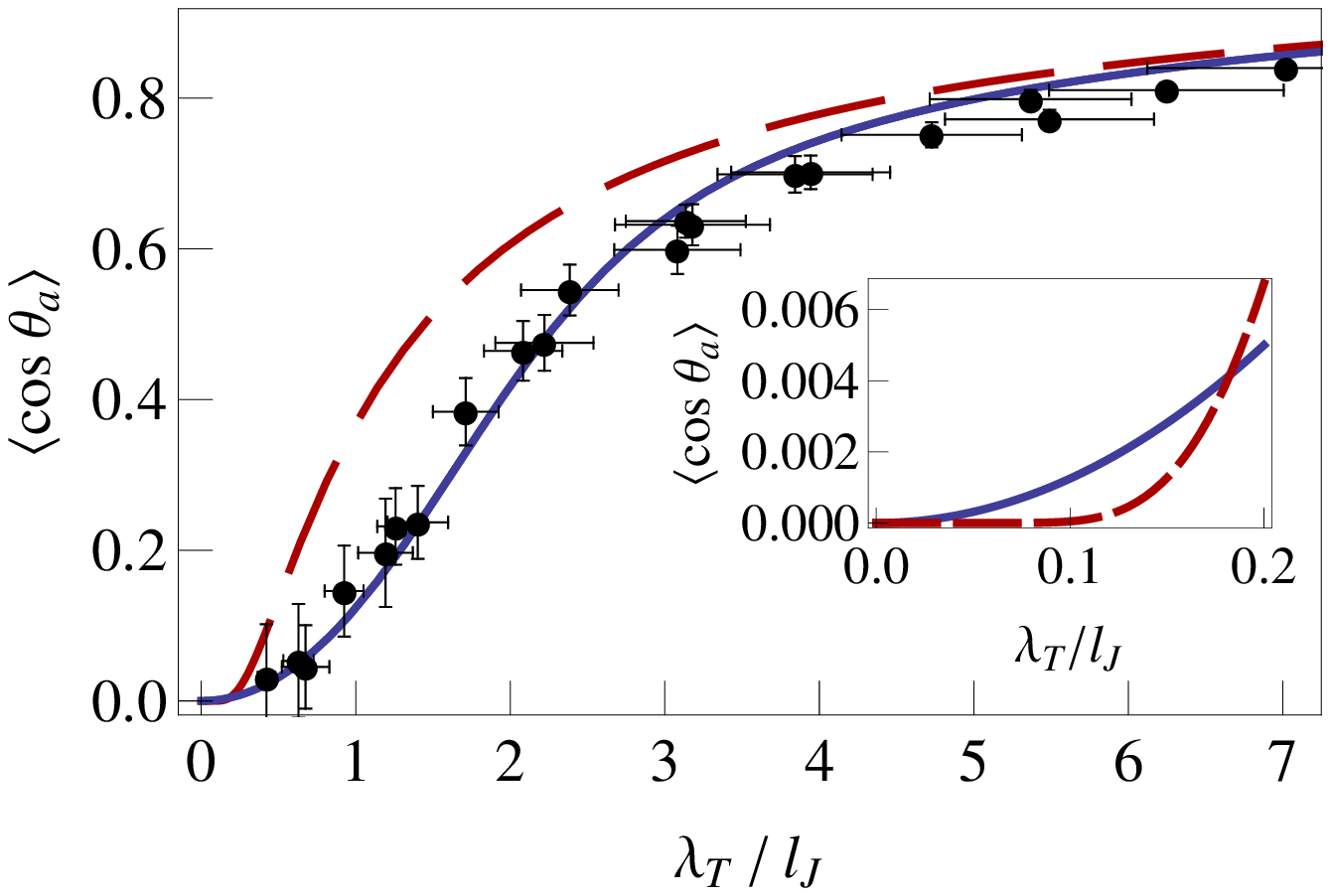,width=0.95\columnwidth} }

\begin{caption} 
{(Color online) Mean interwell contrast  as a function of the ratio of the length scales $\lambda _T$ and $l_J$. Solid line: 
exact theory given by Eq.~(\ref{cosgen}). Dashed line: small-fluctuations approximation $\langle \cos  \theta _\mathrm{a} \rangle = \exp (-l_J/\lambda _T)$ 
following from the linearized theory \cite{WB}. Dots: results of the numerical simulations of the equilibration dynamics of two coupled condensates. 
Units on the axes are dimensionless. Inset: Magnified part of the main plot for small $\lambda _T/l_J$, illustrating the 
high-temperature asymptotics of Eq.~(\ref{cosgen}) in comparison to the linearized theory result. 
\label{F:1}}
\end{caption}
\end{figure} 

To juxtapose the input parameters of our numerical simulations to typical parameters of modern 
atom-chip experiments \cite{Betz,H1,H2}, 
we give the system parameters used in our simulations first in dimensional units, but later show them   
also in dimensionless form. 
The linear density for a single quasicondensate $n_\mathrm{1D}=30~\mu \mathrm{m}^{-1}$ 
and the interaction constant 
$g=2\, \hbar \omega _\perp a_s$ with the radial trapping frequency 
$\omega _\perp =2\pi \times 3$~kHz and the $s$-wave scattering length $a_s=5.3$~nm for $^{87}$Rb yields 
the healing length $\xi \approx 0.35~\mu $m and the Luttinger liquid parameter 
${\cal K} \approx 33$. The periodic boundary conditions were set at an interval of the length $L=100~\mu \mathrm{m}\approx 290 \, \xi $. 
The maximum integration time was $t_\mathrm{max}=0.8$~s. 
After few hundreds milliseconds some kind of equilibrium was obtained. The total energy of the system was conserved in our  
numerical simulations with a good ($\sim 10^{-3}$) accuracy, 
however, it was constantly redistributed in an oscillatory manner between different low-frequency elementary modes, 
including Josephson oscillations. The nonlinear interaction between different modes (see Section~\ref{jxq})  
lead to excitation of Josephson 
oscillations of the total number imbalance $(N_1-N_2)/2$, where $N_j$ is the integral of the density in the $j$th quasicondensate over the whole 
length $L$, i.e., the number of atoms in this quasicondensate, $N_1+N_2\equiv 2N$. In general, the numerical stability of our split-step method was  controlled 
using the criteria of Ref.~\cite{num2}.  The thermal coherence length was 
determined from the phase-correlation functions for each of the two quasicondensates taken \textit{separately} 
by comparison of the numerically obtained value of $\langle \exp [i \theta _j(z)-i\theta _j(z^\prime )]\rangle $, $j=1,2$, 
with its theoretical value $\exp (-|z-z^\prime |/\lambda _T)$ 
for $|z-z^\prime | \lesssim l_J$  \cite{Popov,MoC} (if we trace out the phase and density variables of 
one of the two tunnel-coupled quasicondensates, the properties of its remaining counterpart will be described by the same temperature as of the whole system 
at equilibrium). The averaging is performed over statistically 
uncorrelated (separated by sufficiently large distances) intervals of the whole length $L$ for $|z-z^\prime |\lesssim \lambda _T$. We never obtain complete 
equilibration. In each realization, the correlation length $\lambda _T$ obtained in such a way oscillates around certain mean value, and so does the 
value of $\langle \cos \theta _\mathrm{a} \rangle $ (averaged over the length $L$). Typically, $\lambda _T\approx 8~\mu $m, which corresponds to $T\approx 40$~nK.

We present the results of our numerical simulations in Fig.~\ref{F:1}. Dots represent mean values of $\langle \cos \theta _\mathrm{a} \rangle $ obtained by 
averaging over both the time (on the quasi-equilibration stage of the system evolution) 
and the ensemble of realizations. The error bars in Fig.~\ref{F:1} show  the standard deviations of 
$\langle \cos \theta _\mathrm{a} \rangle $ and $\lambda _T$.  
These error bars indicate 
slow, quasiperiodic variations of $\langle \cos \theta _\mathrm{a} \rangle $ and $\lambda _T$ detected in our simulations. 
The range of $\lambda _T/l_J$ shown in Fig.~\ref{F:1} corresponds to  $J$ increasing from $2\pi \times 0.1$~Hz up to $2\pi \times 8$~Hz. 

To summarize the results of the present Section, we can state that we developed a theory describing the static correlation properties more precisely 
than the harmonic model \cite{WB}. Our approach is based on consideration of the classical partition function for the antisymmetric mode of our problem 
(describable by the sine-Gordon model) and application of the well-known transfer operator technique \cite{Kr1,Kr2}. As one can see from Fig.~\ref{F:1}, the 
difference between our results and those of Ref. \cite{WB} is most apparent for intermediate and small values of $\lambda _T/l_J$ (intermediate 
and weak tunnel coupling).

\section{Josephson oscillations in a noisy extended junction} 
\label{jxq}

The thermal noise effects considered in Sec. \ref{stat} reduce the frequency of Josephson oscillations. 

Consider the absolute number imbalance between two wells, $N_{12}\equiv (N_1-N_2)/2$, 
and its canonically conjugate variable, the overall phase difference $\Phi $ between 
two quasicondensates. In the limit of the atomic repulsion energy dominating over the tunneling, $gn_\mathrm{1D}\equiv gN/L\gg \hbar J$, 
and for small-amplitude oscillations, $|N_1-N_2|\ll N$,  the evolution of these ``global" variables 
is described by the set of equations (see Appendix \ref{A.B}) 
\begin{equation} 
\frac d{dt} \Phi =-\frac {2gN_{12}}{L\hbar }, 
\label{js1} 
\end{equation}
\begin{equation} 
\frac d{dt} N_{12} =  {2J}n_\mathrm{1D} \int _0^L dz\, \sin \theta _a , 
\label{js2} 
\end{equation} 
which is reduced, after elimination of the number-difference variable, to 
\begin{equation} 
\frac {d^2}{dt^2} \Phi = -\omega _\mathrm{J0}^2 \frac {1}{L}\int _0^L dz\, \sin \theta _\mathrm{a} , 
\label{js3}
\end{equation} 
where 
\begin{equation} 
\omega _\mathrm{J0} =\sqrt{4Jgn_\mathrm{1D}/\hbar } 
\label{J0} 
\end{equation} 
is the frequency of the Josephson oscillations for bosonic junction unaffected by thermal noise. At zero temperature, when the thermal noise is absent, and for 
$\ln (L/\xi )\ll {\cal K}$, when the quantum noise can be neglected,  spatial extension of the ultracold-atomic 
Josephson junction plays no role and we can derive Eq. (\ref{J0}) from  the results of Ref. \cite{Smerzi}. 
In the case of small-amplitude Josephson oscillations, the statistical properties of $\cos \theta _\mathrm{a} $ and $\cos (\theta _\mathrm{a}-\Phi )$ do not 
differ significantly, in particular, $\langle \cos \theta _\mathrm{a}\rangle \approx \langle \cos (\theta _\mathrm{a}-\Phi )\rangle $, i.e., the quadratic in $\Phi $ correction is negligible,  
and Eq. (\ref{js3}) reduces to 
\begin{equation} 
\frac {d^2}{dt^2} \Phi +[\omega _\mathrm{J}^2 +\delta \omega _\mathrm{J}^2(t)] \Phi =  \zeta (t) , 
\label{js4}
\end{equation} 
where 
\begin{equation} 
\omega _\mathrm{J}^2=\omega _\mathrm{J0} ^2 \langle \cos \theta _\mathrm{a} \rangle .
\label{OJ2}
\end{equation} 

\begin{figure}[t]  

\centerline{\epsfig{file=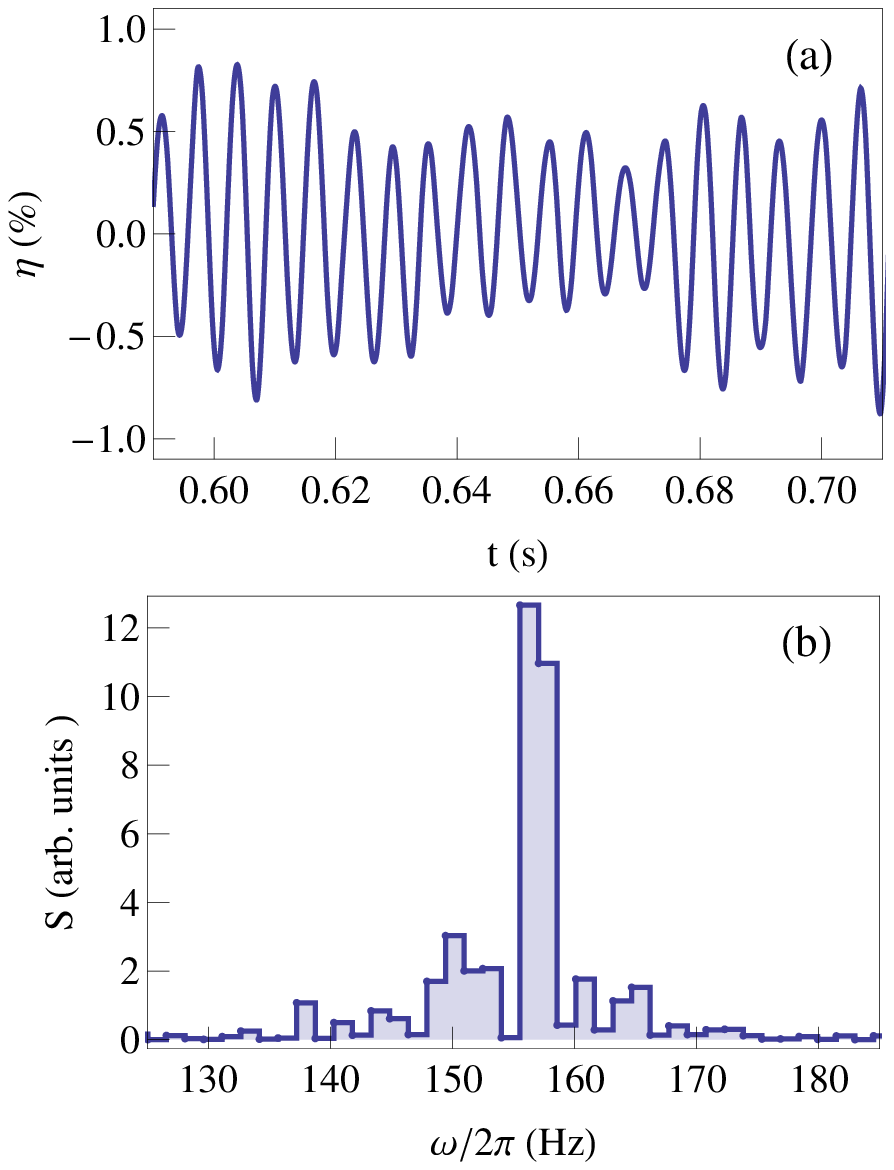,width=0.95\columnwidth }} 

\begin{caption} 
{(Color online) Josephson oscillations for  $J=2\pi \times 8$~Hz  (for other system parameters see Section \ref{stat} of the main text). 
(a) The relative imbalance as a function of time. 
(b) The power spectrum of the atom-number imbalance (averaged over 7 realizations), peaked at theoretically predicted $\omega _\mathrm{J}/(2\pi )=157$ Hz 
and broadened by thermal fluctuations. 
\label{F:2} } 
\end{caption} 
\end{figure} 

\begin{figure}[t]  

\centerline{\epsfig{file=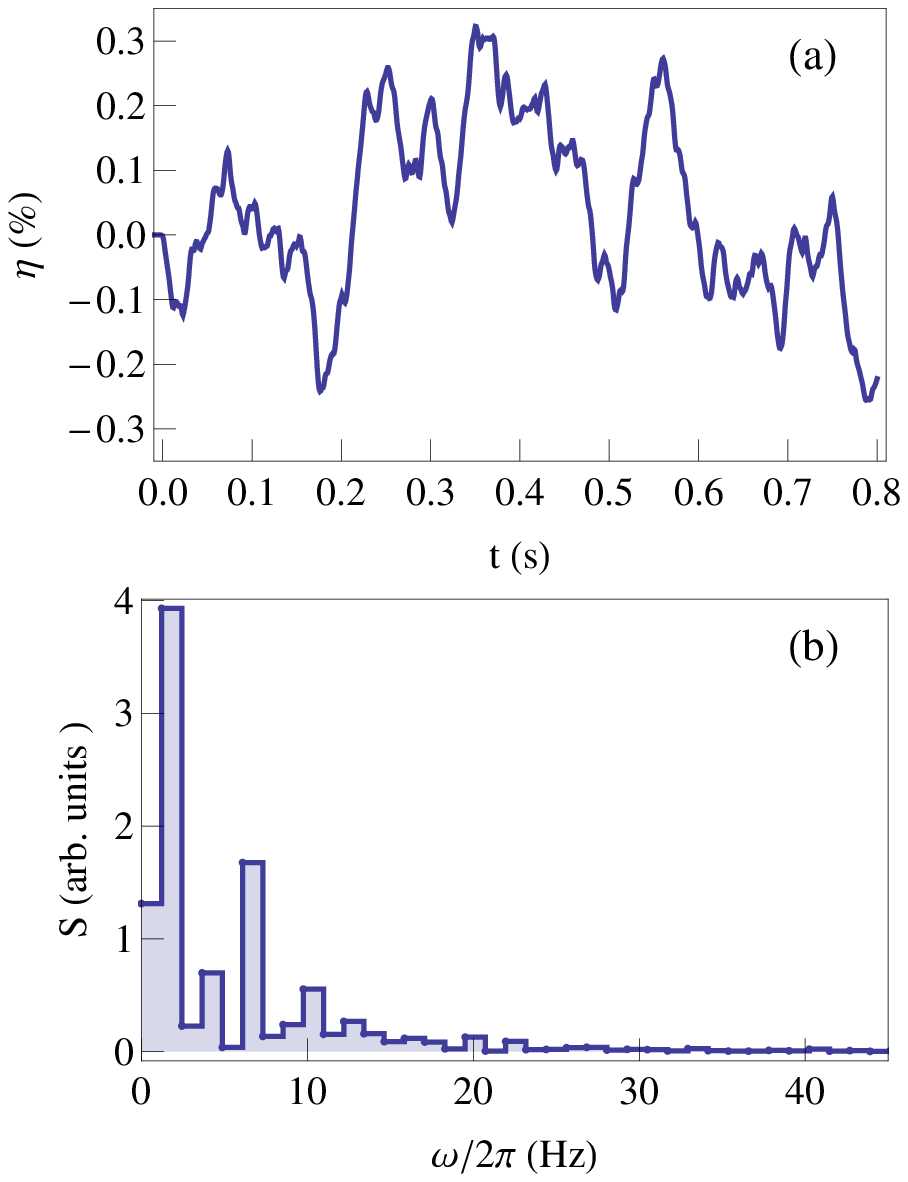,width=0.95\columnwidth }} 

\begin{caption} 
{(Color online) The same as in Fig. \ref{F:2}, but for  $J=2\pi \times 0.1$~Hz (irregular behavior). The spectral peak at 
theoretically predicted $\omega _\mathrm{J}/(2\pi )=6.5$ Hz is smeared out. $S\neq 0$ at $\omega =0$  due to  finite integration time. 
\label{F:3} } 
\end{caption} 
\end{figure} 

In Eq. (\ref{js4}) we explicitly indicate  the time argument of the random driving force 
\begin{equation} 
\zeta (t) = \omega _\mathrm{J0}^2 \frac 1L \int _0^L dz\, \sin (\theta _\mathrm{a}-\Phi ) 
\label{dforce} 
\end{equation} 
and the term 
\begin{equation} 
\delta \omega _\mathrm{J}^2 (t) = \omega _\mathrm{J0}^2 \frac 1L \int _0^L dz\, (\cos \theta _\mathrm{a}-\langle \cos \theta _\mathrm{a} \rangle ) 
\label{ff2} 
\end{equation} 
that describes fluctuations of the oscillation frequency due to the noise of $\theta _\mathrm{a}$ caused by excitations with non-zero momenta. 

\begin{figure}[t]  

\centerline{\epsfig{file=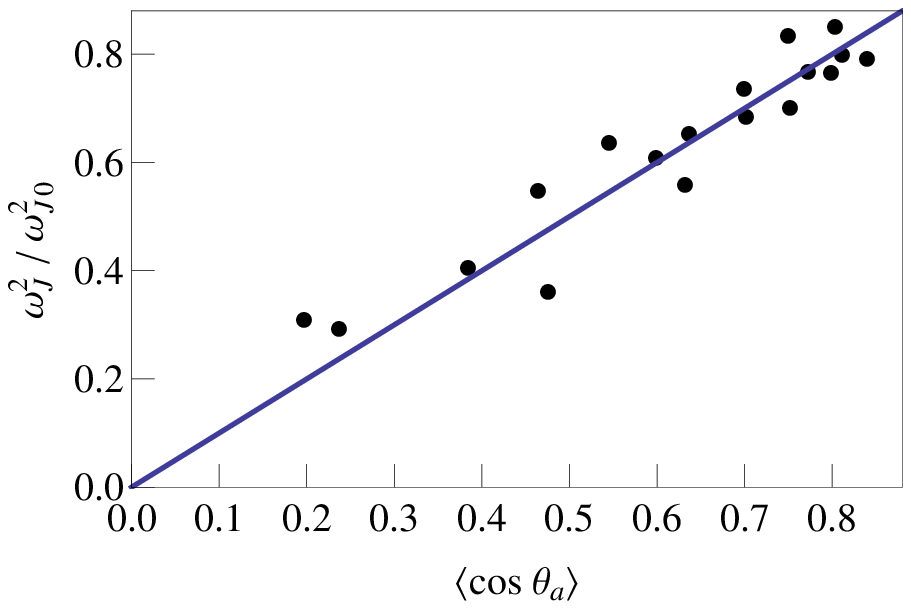,width=0.95\columnwidth }} 

\begin{caption} 
{(Color online) Dots: the square of the frequency of numerically obtained Josephson oscillations (normalized to its value $\omega _\mathrm{J0}^2$ for zero 
thermal noise) as a function of the mean interwell coherence $\langle \cos \theta _\mathrm{a} \rangle $. The straight line: theoretical prediction given by 
Eq.~(\ref{OJ2}). Units on the axes are dimensionless. 
\label{F:4} } 
\end{caption} 
\end{figure} 

If we prepare some appreciable initial imbalance at $t=0$, we obtain, to the first approximation, free Josephson oscillations 
governed by the equation $d^2\Phi /dt^2+\omega _\mathrm{J}^2\Phi =0$, i.e., with the frequency reduced by $\sqrt{\langle \cos \theta _\mathrm{a} \rangle }$ 
compared to the noise-free case of Eq.~(\ref{J0}). 

The presence of the noise broadens the power spectrum of Josephson oscillations 
\begin{equation} 
S(\omega )=\left| \frac 1\tau \int _{t_\mathrm{max}-\tau }^{t_\mathrm{max} }dt\, e^{i\omega t} \eta (t)\right| ^2 
\label{So} 
\end{equation} 
where $\eta =(N_1-N_2)/(2N)$ is the relative number imbalance. The integration in Eq. (\ref{So}) is taken over the time interval $\tau  $  
when the system has already reached its 
nearly-equilibrium state (typically, $\tau \approx 0.65$~s). 
If $\omega _\mathrm{J}$ is high enough, the theory \cite{gitterman} predicts $S(\omega )$ to be a peaked function, 
centered at $\omega _\mathrm{J}$ and having the half-width at the half-maximum of the peak height 
$\gamma =[\hbar L/(8gk_\mathrm{B}T)] \, \mathrm{Re} \int _0^\infty dt^\prime \langle \zeta (t)\zeta (t+t^\prime )\rangle 
\exp (i\omega _\mathrm{J}t^\prime )$. The latter expression, roughly evaluated as 
$\gamma \sim \frac \pi 8 k_\mathrm{B}T/(\hbar {\cal K} \langle \cos \theta _\mathrm{a} \rangle ^2)$, 
correctly describes the order of magnitude of the bandwidth $\Delta \omega /(2\pi )\sim 10$~Hz 
of the numerically obtained spectra $S(\omega )$. 

The presence of the random driving force is the source of excitation of Josephson oscillations in the 
course of the system's evolution, even if initially at $t=0$,  $\Phi =0$ and $\eta  \propto \frac d{dt}\Phi =0$. 
Note, that all the elementary excitations with nonzero momenta in the antisymmetric mode have frequencies larger than than $\omega _\mathrm{J}$. 
The energy transfer between nonzero-momentum excitations and Josephson mode is thus an essentially nonlinear process. The nonlinear structure 
of the right-nand-side of Eq. (\ref{dforce}) provides the presence of the frequency $\omega _\mathrm{J}$ in the spectrum 
$\int _{-\infty }^\infty dt^\prime \langle \zeta (t)\zeta (t+t^\prime )\rangle \exp (i\omega t^\prime )$ of the driving force and thus 
ensures the parametric excitation of the Josephson oscillations. 

We confirmed our analytic estimations by the numerical simulations of two coupled 1D Gross-Pitaevskii equations 
already described in Subsection \ref{subIII.2}. 
An example of a sharp-peaked power spectrum of relative number 
imbalance is given in Fig.~\ref{F:2}, together with an example of time dependence of $\eta $. 

If, on the contrary, $\omega _\mathrm{J} \ll \omega _T$, where $\omega _T=2c/\lambda _T$ is the typical time scale of fluctuations of $\zeta (t)$, 
then the behavior of $\eta (t)$ becomes irregular and $S(\omega )$ does not exhibit a peak at 
$\omega \approx \omega _\mathrm{J}$ any more (see Fig.~\ref{F:3}). 

The results of numerical simulations shown in Figs. \ref{F:2} and \ref{F:3} demonstrate 
certain energy exchange, but no full equilibration between the Josephson oscillations 
and phononic modes. If we set $\Phi \vert _{t=0}=0$ and $\eta \vert _{t=0}=0$ for  $J/(2\pi )=8$~Hz 
(or 0.1~Hz), then at times $t$ between 650~ms and 1~s 
the mean energy of Josephson oscillations is by an order of magnitude (or by 1.5 orders of magnitude, respectively) less than $k_\mathrm{B}T$,  where 
temperature $T$ is determined from the phase-correlation function for a single quasicondensate and is thus associated with the phononic modes. This may 
indicate an extremely long thermalization time for Josephson oscillations. 

We selected our simulations that display a pronounced narrow peak of $S(\omega )$ far from zero frequency (which was the case for $J>2\pi \times 0.7$~Hz), 
estimated  the Josephson frequency $\omega _\mathrm{J}$  and analyzed the dependence of $\omega _\mathrm{J}^2$ on the 
mean interwell coherence. The resulting values are in a good agreement with our theoretical prediction given by Eq.~(\ref{OJ2}), as can be seen from Fig.~\ref{F:4}. 

\section{Conclusion} 
\label{concl}

To conclude, we applied the transfer-operator technique to evaluate coherence and correlation 
properties of two tunnel-coupled 1D weakly-interacting, ultracold systems (quasicondensates) of bosonic atoms. These properties are determined by the ratio of the two 
length scales: $\lambda _T$ that describes the spatial scale of the loss of correlations between two points and $l_J$ that describes the scale for the phase-locking 
between two quasicondensates due to interwell tunneling. In the limit $l_J \lesssim \lambda _T$ the fluctuations of the relative phase are small and we reproduce the 
results of the linearized theory of Ref. \cite{WB}. In the opposite case, we found the mean interwell coherence to decrease much slower ($\propto \lambda _T^2/l_J^2$) 
than the exponential law predicted by the linearized theory. We interprete such a behavior as a signature of thermal creation of sine-Gordon solitons, 
which provide a shift of the relative phase by $2\pi $ and thus do not contribute to the coherence loss, and the corresponding decrease of the 
density of states for phonons (the excitations responsible for the coherence loss at large distances). 

Our analytic estimations are confirmed by numerical modeling of the equilibrium state as a final state of the system's relaxational evolution after a 
quench. This task is solved by extending our numerical method \cite{Grisins} to integration of two coupled 1D Gross-Pitaevskii equations. 

We demonstrate, both analytically and numerically, that thermal fluctuations of the relative phase between two quasicondensates reduce the frequency of Josephson 
oscillations in proportion to $\sqrt{\langle \cos \theta _\mathrm{a} \rangle }$ and broaden their spectrum. If the theoretically predicted value of 
$\omega _\mathrm{J}$ is much less than the bandwidth of the thermal fluctuation (which is of the order of the speed of sound divided by $\lambda _T$), regular 
Josephson oscillations are not observed. 

This work was supported by the the FWF (Project No. P22590-N16). The authors thank T. Berrada and  
J. Schmiedmayer for helpful discussions. 

\appendix

\section{Derivation of Eq. (\ref{cfG})}
\label{A.A} 

We briefly recall here the basics of the transfer operator technique, following Refs. \cite{Kr1,Kr2,transfer1}. 
We introduce a lattice with the step $\Delta z =L/M$, $M$ being the number of sites. We assume cyclic 
boundary conditions, 
\begin{equation} 
\theta _{\mathrm{a}\, M+1}\equiv \theta _{\mathrm{a}\, 1}. 
\label{A1}
\end{equation} 
Then the partition function (\ref{eqZ2}) can  be written as 
\begin{eqnarray} 
Z &=&\int  d\theta _{\mathrm{a}\, 1} \dots \int  d\theta _{\mathrm{a}\, M}\int d\theta _{\mathrm{a}\, M+1}\, 
\delta (\theta _{\mathrm{a}\, M+1}-\theta _{\mathrm{a}\, 1})\times \nonumber \\ && 
\prod _{j=1}^M \exp[ -f(\theta _{\mathrm{a}\, j},\, \theta _{\mathrm{a}\, j+1})] ,
\label{A2}
\end{eqnarray} 
where 
\begin{eqnarray} 
&&f(\theta _{\mathrm{a}\, j},\, \theta _{\mathrm{a}\, j+1})= \frac {\hbar ^2 n_\mathrm{1D}}{4mk_\mathrm{B}T\Delta z} 
\left( \theta _{\mathrm{a}\, j}-\theta _{\mathrm{a}\, j+1}\right) ^2  + \qquad \nonumber \\ && \qquad \qquad 
\frac {\hbar Jn_\mathrm{1D}\Delta z}{k_\mathrm{B}T}(2-\cos \theta _{\mathrm{a}\, j}-\cos \theta _{\mathrm{a}\, j+1})  
\label{A3}  
\end{eqnarray} 
and integrals in our case are taken from $-\pi $ to $\pi $. 
We omit the constant prefactor in Eq. (\ref{A2}) for the sake of simplicity.  
Assume that eigenfunctions $\Psi _n(\theta )$ of the transfer operator 
\begin{equation} 
\int d\theta _{\mathrm{a}\, j}\, e^{ -f(\theta _{\mathrm{a}\, j},\, \theta _{\mathrm{a}\, j+1})} 
\Psi _n(\theta _{\mathrm{a}\, j})=e^{-\epsilon _n\Delta z} \Psi _n(\theta _{\mathrm{a}\, j+1}) 
\label{A4} 
\end{equation} 
form a set, which is complete, orthogonal, and normalized to unity, namely 
\begin{eqnarray} 
\int d\theta \, \Psi _{n^\prime }^* (\theta )\Psi _n(\theta ) &=& \delta _{n^\prime n},        \label{A5} \\
\sum _n \Psi _{n}^* (\theta ^\prime )\Psi _n(\theta ) &=& \delta (\theta ^\prime -\theta ).    \label{A6} 
\end{eqnarray} 
Substituting Eq. (\ref{A6}) into Eq. (\ref{A2}) and using Eq. (\ref{A4}), we obtain 
\begin{equation} 
Z=\sum _n \exp (-\epsilon _nL) .     \label{A7} 
\end{equation} 
The eigenvalues $\epsilon _n$ are positive; in the thermodynamic limit the partition function (\ref{A7}) is 
dominated by the lowest eigenvalue $\epsilon _0$, 
\begin{equation} 
Z\approx \exp (-\epsilon _0L), \qquad L\rightarrow \infty .         \label{A8} 
\end{equation} 

In the continuous limit $\Delta z\rightarrow 0$ Eq. (\ref{A4}) is equivalent to the Schr\"odinger-type 
equation (\ref{aux2}). Strictly speaking, the spectrum of Eq. (\ref{A4}) is shifted with respect to the 
spectrum of Eq. (\ref{aux2}) by a common offset $s_0$, which is related to normalization of the 
eigenfunctions. Since $s_0$ does not depend on $n$, we neglect it in our calculations. 

To calculate correlation functions, in particular, Eq. (\ref{cfG}), we note that $e^{i\theta _\mathrm{a}(z^\prime )}$ and 
$e^{-i\theta _\mathrm{a}(z )}$ act on $\Psi _0$ like quantum-mechanical perturbations, coupling $\Psi _0$ to the 
whole spectrum of eigenfunctions with the matrix elements given by Eq. (\ref{aux1}). Therefore the leading term for 
$\langle \exp [ i \theta _\mathrm{a}(z^\prime ) -
i\theta _\mathrm{a}(z)]\rangle $ in the limit of $L\rightarrow \infty $ is the second-order perturbative correction 
to the propagator for the ground state (with $L$ playing the role of imaginary time), and we obtain thus Eq. (\ref{cfG}).

\section{Derivation of Eqs. (\ref{js1}, \ref{js2})}
\label{A.B} 
  
We begin with the lattice version of the classical sine-Gordon 
Hamiltonian that describes the dynamics of the antisymmetric mode of 
our system: 
\begin{eqnarray} 
H_\mathrm{a}&=&\sum _{j=1}^M 
\Bigg{[}\frac {\hbar ^2n_\mathrm{1D}}{4m\Delta z} \left(  \theta _{\mathrm{a}\, j}-\theta _{\mathrm{a}\, j+1}\right) ^2+
\frac {g}{\Delta z } \delta \! N _{\mathrm{a}\,j}^2 +\nonumber \\ &&
{2\hbar Jn_\mathrm{1D}\Delta z}(1-\cos \theta _{\mathrm{a}\, j})\Bigg{]} ,
\label{B1} 
\end{eqnarray} 
where the $j$th generalized co-ordinate 
$\delta \! N _{\mathrm{a}\,j} =\delta \! \rho _\mathrm{a}\Delta z$ is the half-difference of the atomic numbers in the 1st and 2nd 
quasicondensates at the $j$th site, i.e., the variable canonically conjugate to the local phase difference 
$\theta _{\mathrm{a}\, j}$ (the $j$th generalized momentum). Here we neglect the nonlinear coupling between 
the symmetric and antisymmetric modes, like in Eq. (\ref{Hfullnew}) in the continuous limit.  

For the sake of simplicity, we assume an odd number of sites in the lattice, $M=2M_0+1$, where $M_0$ is a positive integer. 
Then we do a canonical transformation 
\begin{equation} 
\delta \! N _{\mathrm{a}\,j}=\sum _{\ell =-M_0}^{M_0}\delta \! \tilde{N }_{\mathrm{a}}(\ell )\eta (\ell, j) , 
\quad 
\theta _{\mathrm{a}\,j}=\sum _{\ell =-M_0}^{M_0} \tilde{\theta }_{\mathrm{a}}(\ell )\eta (\ell, j) , 
\label{B2}
\end{equation} 
where 
\begin{equation} 
\eta (\ell, j)=\left \{ 
\begin{array}{ll} 
\sqrt{2/M}\cos (2\pi \ell j/M), & \ell =-1,-2, \dots , -M_0 \\
1/\sqrt{M} ,                    & \ell = 0 \\
\sqrt{2/M}\sin (2\pi \ell j/M), & \ell =1,2, \dots , M_0
\end{array}
\right. \!   .  
\label{B3}
\end{equation} 
Then the Hamiltonian (\ref{B1}) reads 
\begin{eqnarray} 
H_\mathrm{a}&=&\sum _{\ell =-M_0}^{M_0} 
\Bigg{\{ }\frac {\hbar ^2n_\mathrm{1D}}{2m\Delta z} \left[  1-\cos (2\pi \ell /M)\right] \tilde{\theta }_{\mathrm{a}}^2(\ell )+
\nonumber \\ && 
\frac {g}{\Delta z } \delta \! \tilde{N }_{\mathrm{a}}^2(\ell )\Bigg{\} } +\label{B4} \\ && 
{2\hbar Jn_\mathrm{1D}\Delta z}\sum _{j=1}^M\left \{ 1-\cos \left[ 
\sum _{\ell =-M_0}^{M_0} \tilde{\theta }_{\mathrm{a}}(\ell )\eta (\ell, j)\right] \right \}   . \nonumber  
\end{eqnarray}
From the Hamiltonian equations 
\begin{equation} 
\frac d{dt}\delta \! \tilde{N}_{\mathrm{a}}(\ell )=\frac {\partial H_\mathrm{a}}{\partial \, \tilde{\theta }_{\mathrm{a}}(\ell )},
\quad 
\frac d{dt}\tilde{\theta }_{\mathrm{a}}(\ell )=-\frac {\partial H_\mathrm{a}}{\partial \, \delta \! \tilde{N}_{\mathrm{a}}(\ell )}
\label{B5} 
\end{equation} 
we find, in particular, 
\begin{equation} 
\frac d{dt}\tilde{\theta }_{\mathrm{a}}(0)=-\frac {2g\delta \! \tilde{N}_{\mathrm{a}}(0)}{\Delta z} , 
\label{B6} 
\end{equation}  
\begin{equation} 
\frac d{dt}\delta \! \tilde{N}_{\mathrm{a}}(0)= {2\hbar Jn_\mathrm{1D}\Delta z}\sum _{j=1}^M\sin \left[ 
\sum _{\ell =-M_0}^{M_0} \tilde{\theta }_{\mathrm{a}}(\ell )\eta (\ell, j)\right] \! .\quad 
\label{B7}  
\end{equation} 
In the limit of $\Delta z\rightarrow 0$ the sums over $j$ converge to integrals over $z$. Taking into account that 
$N_{12}=\int dz\, \delta \! \rho _\mathrm{a}=\sum _{j=1}^M \delta \! N _{\mathrm{a}\,j}=\sqrt{M} 
\delta \! \tilde{N}_{\mathrm{a}}(0)$, identifying the generalized momentum conjugate to $N_{12}$ as 
$\Phi = \tilde{\theta }_{\mathrm{a}}(0)/\sqrt{M}=(1/M)\sum _{j=1}^M \theta _{\mathrm{a}\,j}$ and recalling that 
$L=M\Delta z$, we obtain Eqs. (\ref{js1},~\ref{js2}). The spatially fluctuating part of the phase is then 
$\theta _\mathrm{a}-\Phi $.

\end{document}